\documentclass[aps,prd,twocolumn,amsmath,amssymb,floatfix,nofootinbib,superscriptaddress]{revtex4-1}

\newcommand{\nhat}{\hat{ \mathbf{n}}}

\usepackage{graphicx}
\usepackage{dcolumn}
\usepackage{bm}
\usepackage[usenames, dvipsnames]{color}
\usepackage[normalem]{ulem}

\graphicspath{ {figures/} }

\usepackage{graphicx}
\usepackage{amssymb}
\usepackage{mhchem}

\usepackage{epstopdf}
\usepackage{graphicx}
\usepackage{amssymb}
\usepackage{epstopdf}
\usepackage{graphicx}
\usepackage{dcolumn}
\usepackage{bm}
\usepackage{color}
\usepackage{mhchem}

\newcommand{\be}{\begin{eqnarray}}

\newcommand{\ee}{\end{eqnarray}}

\newcommand{\bk}{\mathbf{k}}
\newcommand{\khat}{\hat{\mathbf{k}}}

\usepackage[export]{adjustbox}

\def\vl{\boldsymbol{\ell}}
\def\vlp{\boldsymbol{\ell}'}
\def\vL{\mathbf{L}}
\def\vLp{\mathbf{L}'}

\def\nhat{\hat{\mathbf{n}}}

\newcommand{\apjl}{Astrophys. J. Lett.}

\newcommand{\dd}{{\rm d}}

\begin{document}

\title{Beyond CMB cosmic variance limits on reionization with the polarized SZ effect}
\newcommand{\cita}{Canadian Institute for Theoretical Astrophysics, University of Toronto, 60 St.~George Street, Toronto, Canada, M5S 3H8}
\newcommand{\kavli}{Kavli Institute for Cosmology, Madingley Road, Cambridge, UK, CB3 0HA}
\newcommand{\damtp}{DAMTP, Centre for Mathematical Sciences, Wilberforce Road, Cambridge, UK, CB3 0WA}
\newcommand{\princeton}{Department of Astrophysical Sciences, Princeton University, Princeton, NJ, USA, 08544}
\newcommand{\cornell}{Department of Astronomy, Cornell University, Ithaca, NY, USA, 14853}
\newcommand{\cca}{Center for Computational Astrophysics, Flatiron Institute, 162 Fifth Avenue, New York, NY, USA, 10010}

\author{Joel Meyers}
\affiliation{\cita}

\author{P. Daniel Meerburg}
\affiliation{\cita}
\affiliation{\kavli}
\affiliation{\damtp}
\author{Alexander van Engelen}
\affiliation{\cita}

\author{Nicholas Battaglia}
\affiliation{\cca}
\affiliation{\princeton}
\affiliation{\cornell}

\date{\today}

\begin{abstract}

Upcoming cosmic microwave background (CMB) surveys will soon make the first detection of the polarized Sunyaev-Zel'dovich effect, the linear polarization generated by the scattering of CMB photons on the free electrons present in collapsed objects. Measurement of this polarization along with knowledge of the electron density of the objects allows a determination of the quadrupolar temperature anisotropy of the CMB as viewed from the space-time location of the objects.  Maps of these remote temperature quadrupoles have several cosmological applications.  Here we propose a new application: reconstruction of the cosmological reionization history.  We show that with quadrupole measurements out to redshift 3, constraints on the mean optical depth can be improved by an order of magnitude beyond the CMB cosmic variance limit.

\end{abstract}

\maketitle


\section{Introduction}

The cosmological reionization history is a target of considerable scientific interest and also a source of uncertainty for interpreting measurements of the cosmic microwave background (CMB) anisotropies. Reionization is perhaps the only time that luminous sources directly and drastically altered the state of the entire Universe, as they converted the cold and neutral intergalactic medium to a warm and highly ionized one. Current knowledge of the reionization history is very limited. In addition to quasar absorption \citep{Fan:2006dp} and 21cm \citep{Monsalve:2017mli} measurements, primary and secondary CMB anisotropies observations have begun to place constraints on the timing and duration of the reionization.
The current best constraint on the optical depth comes from an analysis of the data obtained with the \textit{Planck} satellite, giving $\tau = 0.058 \pm 0.012$ \citep{Adam:2016hgk}. The South Pole Telescope team found evidence for the kinetic Sunyaev-Zel'dovich  angular power spectrum at $2 \sigma$ \citep{George:2014oba}, which when combined with the mean optical depth measurement, implies a median reionization redshift of $7.8$ and duration less than $2.8$ in redshift (95\% upper limit)~\citep{Adam:2016hgk}.

One of the primary scientific goals of next-generation cosmological surveys is an absolute measurement of the sum of neutrino masses. This measurement is made by comparing the primordial amplitude of density fluctuations with the amplitude at recent times. In order to determine the primordial amplitude from observations of CMB anisotropies, one must break the existing degeneracy with the mean optical depth due to reionization, $\tau$~\cite{Zaldarriaga:1997ch,Allison:2015qca}. 
Given the planned configurations of upcoming CMB experiments and large scale structure surveys, an error on the optical depth of about $\sigma({\tau}) \lesssim 0.005$ is required in order to achieve a $3\sigma$ detection of the sum of neutrino masses in the case of a normal hierarchy with a single nearly massless eigenstate, i.e., 
$\sigma\left({\sum m_{\nu}}\right) \leq 0.02$~eV~\cite{CMBS4}.  

The degeneracy between the primordial amplitude and the optical depth can be broken with a measurement of large-scale $E$-mode CMB polarization, which is generated during reionization and as such is directly proportional to the optical depth~\cite{Zaldarriaga:1996ke}.  
Direct measurements of large scale $E$-mode polarization is challenging due to astrophysical foregrounds and instrumental systematics, and observations from the ground face additional complications associated with the atmosphere.
Proposed observations from space~\cite{Kogut:2011xw,Matsumura:2013aja,2015hsa8.conf..334R} and ongoing observations from the ground~\cite{CLASSexp} aim to improve the measurements of large scale CMB polarization measurements. Alternative measurements have been proposed to constrain the optical depth \cite{21cmTau,Meerburg:2017lfh}, but these proposals face their own limitations and systematic uncertainties, and it remains to be seen if these methods will prove successful. Besides, even with perfect measurements of CMB polarization, our ability to constrain the optical depth with large scale $E$ modes is hampered by the limited number of independent modes, a problem which is exacerbated when only a fraction of the sky can be sufficiently cleaned.
The error achievable with cosmic-variance-limited measurements of large scale CMB polarization on the whole sky is $\sigma(\tau) \simeq 0.002$~\cite{CMBS4}.

In this paper we propose a method whereby one can circumvent the limitations imposed by sample variance in order to better constrain the reionization history, and thus the mean optical depth $\tau$. CMB polarization is generated by the Thomson scattering of CMB photons by free electrons in the presence of quadrupolar anisotropy of the incoming flux~\cite{Rees:1968abc,Zaldarriaga:1996xe}. Typical analyses assume only statistical knowledge of the fluctuations responsible for generating quadrupolar anisotropy. However, direct measurements of remote quadrupoles can be achieved by measuring the polarized Sunyaev-Zel'dovich (pSZ) effect \cite{Kamionkowski:1997na}. Free electrons present in collapsed objects such as galaxies and galaxy clusters scatter CMB photons, inducing a linear polarization which is proportional to the projected quadrupolar temperature anisotropy of the CMB as viewed from the space-time location of the object~\cite{Zeldovich:1980abc,Sunyaev:1980nv,Sunyaev:1981abc,Sazonov:1999zp}. 
Thus, measuring the polarization in the direction of the object allows a determination of two components of the remote temperature quadrupole if the Thomson optical depth of the object can be inferred independently. The field of remote quadrupoles has a very large coherence length, implying that remote quadrupoles at relatively low redshifts are correlated with the temperature quadrupoles present at reionization. Cross-correlating maps of remote temperature quadrupoles with CMB $E$-mode polarization on large scales thereby allows for improved constraints on the free electron density as a function of redshift.


\begin{figure}
	\includegraphics[width = \columnwidth]{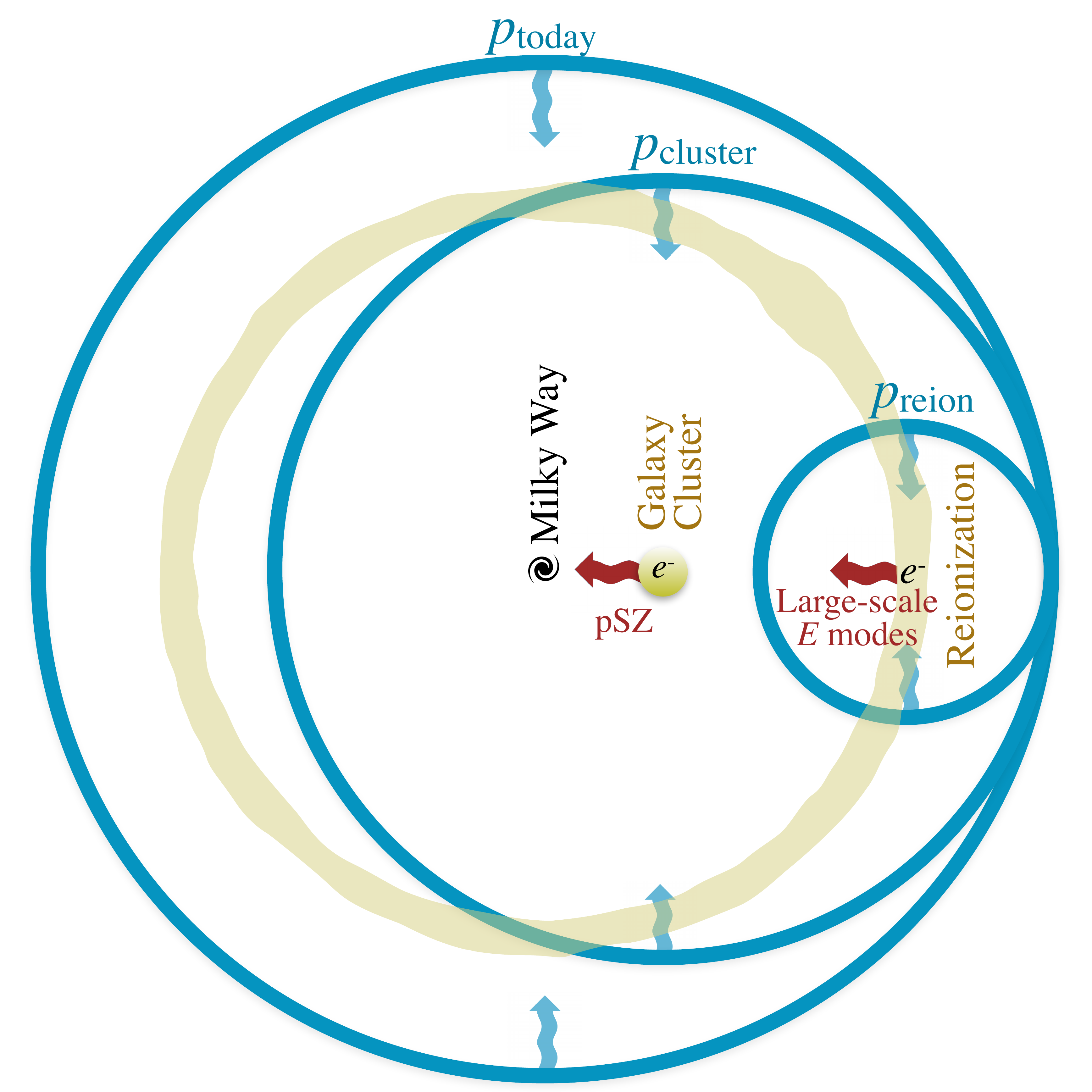}
	\caption{The CMB polarization seen in the direction of a collapsed object such as a galaxy cluster is proportional to the optical depth due to free electrons in the object and the incident temperature quadrupole seen from the space-time location of the object, $p_\mathrm{cluster}$.  The pSZ effect can be used to measure $p_\mathrm{cluster}$, which is correlated with the quadrupole field incident on free electrons during reionization, $p_\mathrm{reion}$. The latter field is in turn responsible for generating large-scale $E$ modes in the CMB.  We can reduce cosmic variance on the reionization history by taking advantage of this correlation.} 
    \label{fig:spectra}
\end{figure}

Previous studies have shown that remote quadrupole measurements could be used to achieve independent measurements of large scale fluctuations, thereby lowering cosmic variance on these modes~\cite{Kamionkowski:1997na} which provides a potentially useful probe of large scale CMB anomalies~\cite{Cooray:2002cb,Seto:2005de}. Observing the redshift evolution of remote quadrupoles can also be used as a probe of late-time structure formation and thus dark energy~\cite{Cooray:2002cb,Cooray:2003hd}. Turning things around, we can also use our knowledge of the local temperature quadrupole in combination with the pSZ signal generated in nearby objects to determine the optical depth, and thus the baryon content, of the objects~\cite{Louis:2017hoh}.  
The statistics of remote quadrupoles have been well studied analytically~\cite{Portsmouth:2004mk,Bunn:2006mp} and in simulations~\cite{Amblard:2004yp,Ramos:2012abc}. The pSZ effect has not yet been observed, but it should be detectable with upcoming CMB and galaxy surveys ~\cite{Hall:2014wna,Deutsch:2017cja,Deutsch:2017ybc}, encouraging further exploration of this observable as a probe of cosmology.


\section{Remote Quadrupoles}

In this section, we review the formalism and statistics of remote quadrupoles following Refs.~\cite{Bunn:2006mp,Hall:2014wna}.   
The complex CMB polarization along a direction $\nhat$ can be written as~\cite{Hu:1997hp}
\be \label{eq:CMBpol}
	(Q\pm iU)(\nhat) =  \int_{\eta_0}^{\eta_\star} \dd{\eta} \, \dot{\tau}(\eta) e^{-\tau(\eta)} p^{\pm}(\nhat, \eta), 
\ee
where $\eta$ is the conformal time, with $\eta_0$ today, and $\eta_\star$ the conformal time of last scattering. The polarization source function is given by 
\be \label{eq:polSource}
    p^{\pm}(\nhat, \eta) = - \frac{\sqrt{6}}{10} \sum_m {}_{\mp 2} Y_{2 m}(\nhat) T_{2m}(\nhat,\eta) \, ,
\ee
where $T_{2m}(\nhat,\eta)$ is the temperature quadrupole\footnote{Note that when the incident radiation is polarized, $T_{2m}$ in this expression should be replaced by $T_{2m} - \sqrt{6}E_{2m}$, where $E_{2m}$ is the $E$-mode quadrupole~\cite{Hu:1997hp,Hall:2014wna}.  This is a very small correction at late times, but we include it in our numerical analysis.} at conformal time  $\eta$.  The polarization can be expanded in spin-2 spherical harmonics as~\cite{Chon:2003gx}
\be \label{eq:EBlm}
	(Q\pm iU)(\nhat) = \sum_{\ell m} (E_{\ell m} \mp i B_{\ell m}) {}_{\mp2}Y_{\ell m}(\nhat) \, .
\ee
Similarly, the polarization due to scattering in an object of optical depth $\delta\tau$ at redshift $z$ along direction $\nhat$ is given by~\cite{Bunn:2006mp,Hall:2014wna}
\be \label{eq:clusterPol}
	(Q\pm iU)^{(\delta\tau)}(\nhat) =  \delta\tau(\nhat,\eta(z)) p^{\pm}(\nhat, \eta) \, , 
\ee
which can be expanded in spin-2 spherical harmonics as
\be \label{eq:plm}
	&& (Q\pm iU)^{(\delta\tau)}(\nhat) \delta\tau^{-1}(\nhat,\eta(z)) = \nonumber \\
    && \qquad \sum_{\ell m} \left(p^E_{\ell m}(\eta) \mp i p^B_{\ell m}(\eta)\right) {}_{\mp 2}Y_{\ell m}(\nhat) \, .
\ee
In the presence of purely scalar fluctuations the CMB polarization produced by scattering on the mean density of free electrons contains only $E$ modes ($B_{\ell m} = 0$), and the remote quadrupoles have the parity properties of $E$-mode polarization ($p^B_{\ell m} = 0$). Note, however that polarization resulting from the pSZ effect contains both $E$ and $B$ modes, due to the spatial variations of the optical depth induced by the objects~\cite{Hu:1999vq,Dvorkin:2009ah,Alizadeh:2012vy}. We will assume vanishing tensor fluctuations and drop the $E$ superscript on $p_{\ell m}$ from here on. With this assumption, the coefficients are given by
\be \label{eq:plmMode}
	p_{\ell m} (\eta) &=& -i^{\ell} 3\pi \sqrt{\frac{(\ell+2)!}{(\ell-2)!}} \nonumber \\ 
    && \times \int \frac{\mathrm{d}^3\bk}{(2\pi)^{3/2}} \frac{j_\ell(k\eta)}{(k\eta)^2} \Delta_2(k,\eta) \Phi(\bk) Y_{\ell m}^*(\khat) \, .
\ee
Here, $\Phi$ is the three-dimensional gravitational potential of the Universe, $\Delta_2$ is the cosmological transfer function for $\ell = 2$ quadrupole anisotropies, and $j_\ell$ is the spherical Bessel function.


\begin{figure}
	\includegraphics[width = \columnwidth]{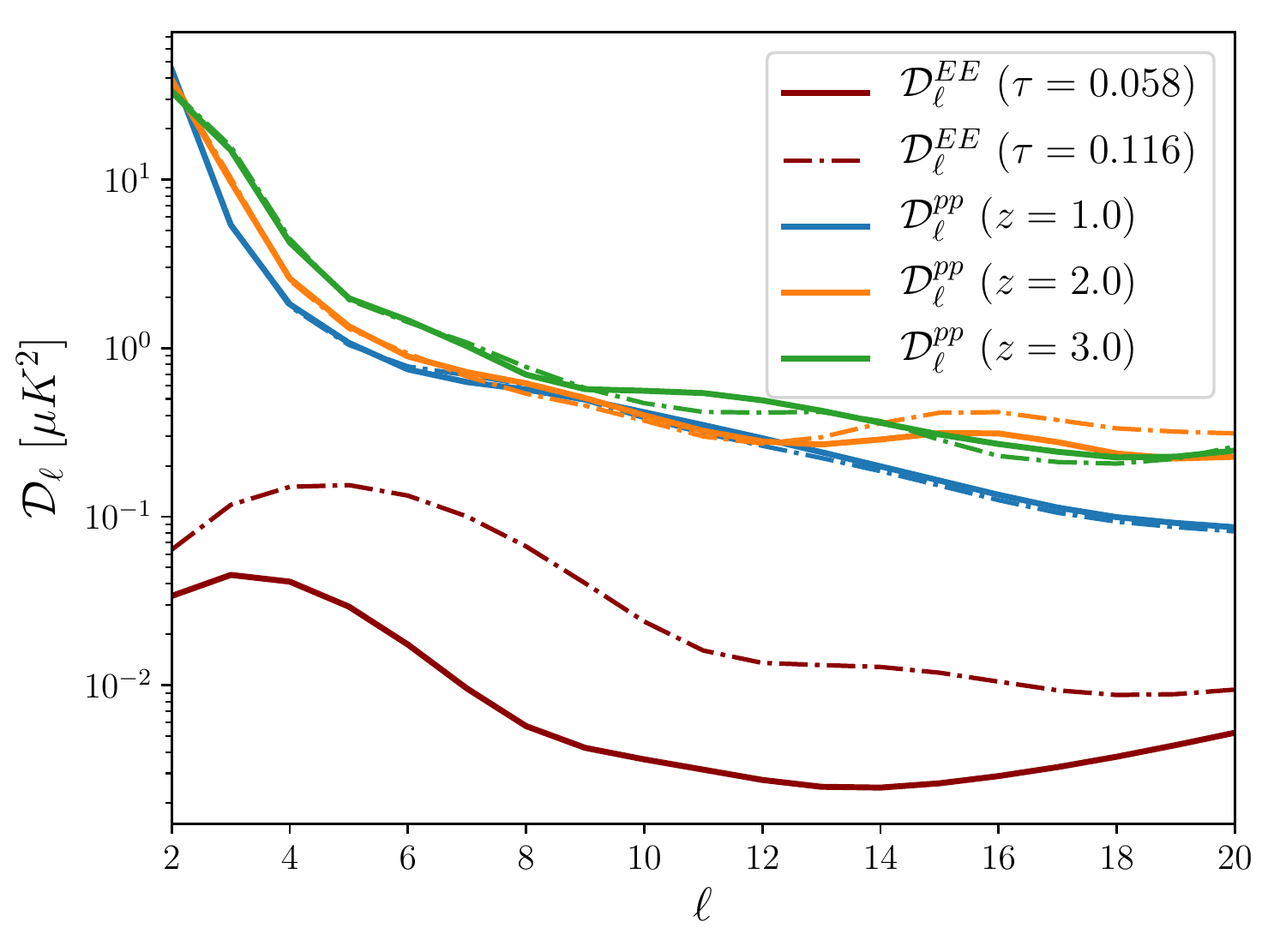}
	\caption{Power spectrum of $E$-mode polarization along with the remote quadrupole autospectrum at three redshifts, each for two values of the mean optical depth. Here we have defined $\mathcal{D}_\ell \equiv \ell(\ell+1) C_\ell / 2\pi $. } 
    \label{fig:spectra}
\end{figure}

We can now compute all of the relevant auto- and cross-spectra including the remote temperature quadrupoles as well as the CMB temperature and $E$-mode polarization
\be \label{eq:spectra}
	\left\langle X_{\ell m} Y_{\ell' m'}^* \right\rangle = \delta_{\ell \ell'}\delta_{m m'} C_\ell^{XY} \, ,
\ee
where $X,Y \in \{T,E,p(\eta)\}$.  At low redshift, the remote quadrupole is strongly correlated with the observed CMB temperature quadrupole, which has already been well measured; it thus does not provide new cosmological information~\cite{Bunn:2006mp,Hall:2014wna}. However, at higher redshifts, it is correlated with CMB $E$-mode polarization generated during reionization, making $C_\ell^{Ep(\eta)}$  sensitive to the reionization history.

Since the polarization is sourced by the field of  remote temperature quadrupoles $p(\eta)$, the harmonic coefficients of the CMB $E$ modes can be written as 
\be \label{eq:CMBplm}
	E_{\ell m}  = \int_{\eta_0}^{\eta_\star} \! \dd{\eta} \, g(\eta) p_{\ell m}(\eta) \, ,
\ee
where we have defined the visibility function
\be \label{eq:visibility}
	g(\eta) =  \dot{\tau}(\eta) e^{-\tau(\eta)} \, ,
\ee
which gives the probability that a CMB photon last scattered at conformal time $\eta$ for a spatially homogeneous universe. It is clear from Eq.~(\ref{eq:CMBplm}) that given sufficiently sensitive measurements of the realizations of both CMB $E$ modes and $p(\eta)$, there should be no cosmic variance limit on $g(\eta)$, or the total optical depth $\tau$.
This proposal is very similar to combining observations of two tracers of large scale structure in order to cancel cosmic variance on measurements of the halo bias~\cite{Seljak:2008xr}. 
We will combine three-dimensional measurements of $p(\eta)$ with polarization maps.  For each distance $\eta_i$ where we have a measurement of the remote quadrupole field $p(\eta_i)$, we will obtain some information about the visibility function over a range of distances centered on $\eta_i$. We can combine these measurements to constrain the reionization history.  The large coherence length of the remote quadrupole field allows measurements of $p(\eta)$ even at  redshifts much lower than the end of reionization, $z \sim 6$, to provide useful information about reionization.

In addition to the cosmic variance cancellation, measurements of remote temperature quadrupoles provide a very small amount of direct information regarding the reionization history.  The effect of reionization on the CMB temperature is to suppress anisotropies on scales smaller than the horizon at the redshift of reionization and to produce some additional fluctuations around the same scale due to Doppler scattering~\cite{Sugiyama:1993dq,Hu:1995fqa}.
As can be seen in Fig.~\ref{fig:spectra}, this dependence is weak, and the constraining power demonstrated in the next section will be dominated by the cancellation of cosmic variance described above.


\section{Forecasts}

We now turn to what can be achieved with upcoming experiments.  First we will discuss the ability of upcoming surveys to observe remote temperature quadrupoles through measurement of the pSZ effect. We then discuss how maps of CMB polarization and three-dimensional maps of the remote quadrupole field can be combined to improve constraints on the cosmological reionization history.

\subsection{Observing Remote Quadrupoles}


\begin{figure}
	\includegraphics[width = \columnwidth]{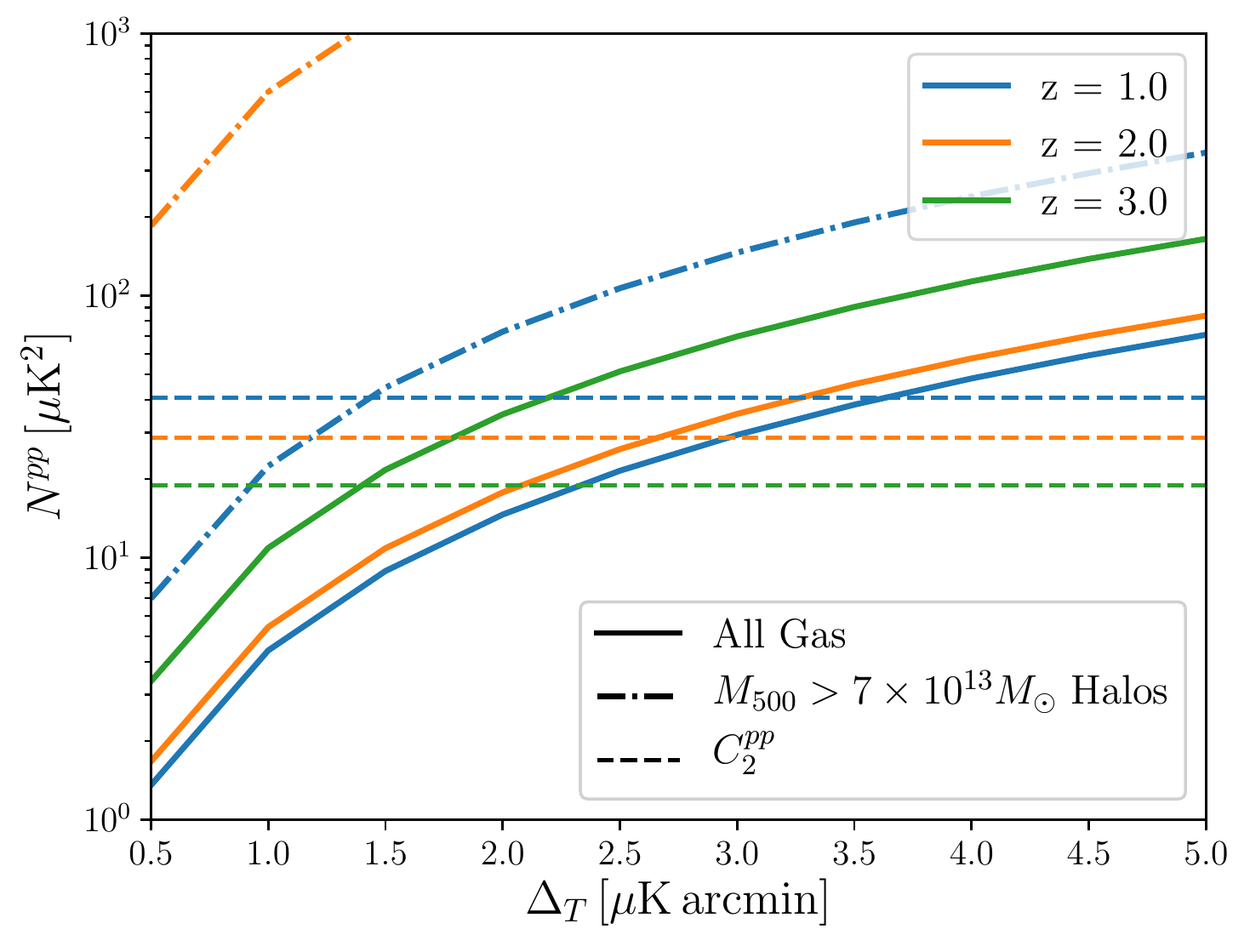}
	\caption{Noise on the reconstructed remote quadrupole field for various redshifts as a function of CMB noise assuming a 1-arcminute beam for the CMB survey and for two assumptions about the independently inferred free electron density: using all of the gas (solid) and using gas in halos of mass $M_{500} > 7\times 10^{13}\, M_\odot$ (dash-dot).  In both cases, the gas is split into bins of width $\Delta z = 0.5$, and reconstruction is performed in each bin, though we show the noise only at integer redshifts for clarity. We also plot the power $C_\ell^{pp}$ predicted at $\ell=2$ for the same set of redshifts. Note that for the CMB-S4 noise level of about 1 $\mu$K-arcmin, the remote quadrupole field at $\ell=2$ can be mapped at low redshift.} \label{fig:pp_noise_CMB}
\end{figure}

With a given catalog of objects at known redshifts and with independently inferred optical depth profiles, one could use an observed map of CMB polarization to make a local estimate of the remote quadrupole field at the location of each object.  In the limit of many objects spread over the sky this procedure becomes a quadratic estimator for the spatially varying remote quadrupole field  between the observed CMB polarization and the independently inferred optical depth fluctuations, similar to those performed on the CMB to estimate gravitational lensing~\cite{Hu:2001kj}. In the Appendix, we derive this minimum variance quadratic  estimator and show that the variance on the reconstructed field of remote quadrupoles on large scales is very nearly white noise given by
\be \label{eq:reconstruction_noise}
    N^{pp(\eta)}_L &=& \left[\int \! \frac{\mathrm{d}^2{\vl}}{(2\pi)^2} \, \rho_{|\vl-\vL|}^2(\eta) \, C_{|\vl-\vL|}^{\delta\tau\delta\tau(\eta)} \right. \\
    && \quad \times \left(C_{\ell}^{EE,\mathrm{obs}}\cos^2(2(\phi_{\vL} - \phi_{\vl})) \right. \nonumber \\  
    &&  \qquad \left. \left. + C_{\ell}^{BB,\mathrm{obs}}\sin^2(2(\phi_{\vL}-\phi_{\vl}))\right)^{-1} \right]^{-1} \, , \nonumber
\ee
where $\rho_\ell(\eta)$ is the cross correlation coefficient between the independently inferred map of $\delta \tau(\eta)$ and the actual optical depth fluctuations at that conformal time, $C_\ell^\mathrm{obs}$ refers to the observed power spectrum including noise, and $\phi_{\vl}$ is the angle that the wavenumber ${\vl}$ makes with respect to the $x$-axis.


\begin{figure}[t]
	\includegraphics[width = \columnwidth]{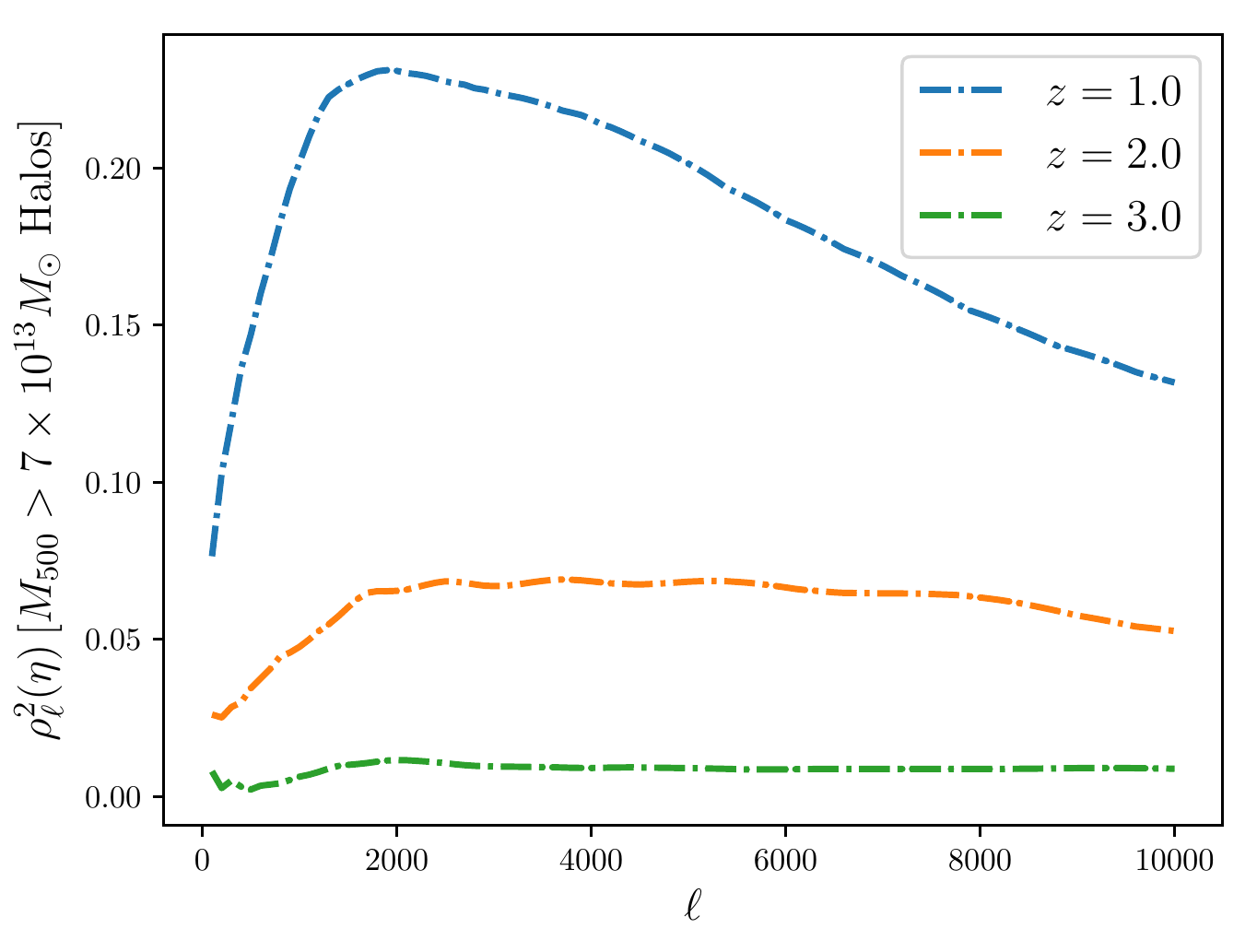}
	\caption{The correlation coefficient $\rho_\ell^2(\eta)$ of the inferred and true $C_\ell^{\delta\tau \delta\tau(\eta)}$ assuming only observation of clusters which will be detected in a futuristic CMB survey, namely those in halos with $M_{500}>7\times 10^{13}M_\odot$.  This correlation coefficient could be increased by including measurements of lower mass objects detected in a galaxy survey, for example.} \label{fig:rho_halo}
\end{figure}

In Fig.~\ref{fig:pp_noise_CMB} we show the noise $N^{pp(\eta)}_L$ for a set of CMB experiments and for two simulation-based treatments of the inferred electron density.  We consider CMB surveys with a 1-arcminute beam across a range of noise levels, and assume that $C_\ell^{EE}$ and $C_\ell^{BB}$ have both been delensed using an internal estimate of the lensing field~\cite{Knox:2002pe,Kesden:2002ku,Hirata:2003ka,Smith:2010gu,Green:2016cjr} which helps to reduce the variance by reducing the power of small scale polarization.  We use the simulations of Ref.~\cite{BBPSS2010} to calculate $C_\ell^{\delta\tau\delta\tau(\eta)}$, and show the noise obtained assuming an optimistic case of a measurement of all of the ionized gas ($\rho_\ell(\eta) = 1$). The same simulations are used to isolate the electrons in halos that would be detected in a futuristic CMB survey ($\rho_\ell(\eta) < 1$), namely those representing galaxy clusters  exceeding a mass of $M_{500} = 7\times 10^{13}\, M_\odot$ (see Fig.~\ref{fig:rho_halo}).  Here, $M_{500}$ is the spherical overdensity mass with respect to the critical density of the Universe. These 
simulations were run using GADGET-2 \citep{Gadget2} and include sub-grid models for radiative cooling, star formation, supernova feedback, and AGN feedback. Halos in these simulations were first found using a friends-of-friends algorithm~\citep{FOF}; then, for every halo, new centers of mass were computed iteratively followed by new spherical overdensity masses.  The noise curves shown in Fig.~\ref{fig:pp_noise_CMB} depend on how we choose to bin $\delta\tau(\eta)$, and we take bins of fixed width in redshift space with $\Delta z = 0.5$.

\subsection{Constraining Reionization}

We now wish to determine how including measurements of remote quadrupoles impacts cosmological constraints.
After cleaning Galactic foregrounds, the observed $E$-mode polarization of the CMB on large angular scales is dominated by the polarization generated by scattering of remote temperature quadrupoles with the mean electron density during reionization. Restricting ourselves to those scales ($\ell\lesssim 20$) the $E$-mode power spectrum is given by
\be \label{eq:ClEE}
    C_\ell^{EE} = \int_{\eta_0}^{\eta_\star} \! \dd{\eta} \int_{\eta_0}^{\eta_\star} \! \dd{\eta'} \, g(\eta)g(\eta') C_\ell^{p(\eta)p(\eta')} \, .
\ee
and the cross spectrum with the remote quadrupole field is given by
\be \label{eq:ClEp}
    C_\ell^{Ep(\eta')} = \int_{\eta_0}^{\eta_\star} \! \dd{\eta} \, g(\eta) C_\ell^{p(\eta)p(\eta')} \, .
\ee

Constraints on reionization can be obtained by computing the auto- and cross-spectra of given maps of $E$ and $p(\eta)$ and taking account of the covariance among them.
We can forecast the constraints obtained from such a procedure by computing the Fisher matrix defined as \cite{Fisher1935,Knox:1995dq,Jungman:1995bz}
\be
	F_{ij} = \sum_\ell \frac{2\ell + 1}{2} f_\mathrm{sky} \mathrm{Tr} \left(\mathbf{C}_\ell^{-1} \frac{\partial \mathbf{C}_\ell}{\partial \lambda_i} \mathbf{C}_\ell^{-1} \frac{\partial \mathbf{C}_\ell}{\partial \lambda_j} \right) \, ,
\ee
with the covariance matrix $\mathbf{C}_\ell$ given by
\be 
	\mathbf{C}_\ell = 
    \begin{bmatrix}
    C_\ell^{TT} & C_\ell^{TE} &  \dots  & C_\ell^{Tp(\eta_n)} \\
    C_\ell^{TE} & C_\ell^{EE} &  \dots  & C_\ell^{Ep(\eta_n)} \\
    C_\ell^{Tp(\eta_1)} & C_\ell^{Ep(\eta_1)} & \dots & C_\ell^{p(\eta_1)p(\eta_n)} \\
    \vdots & \vdots & \ddots & \vdots \\
    C_\ell^{Tp(\eta_n)} & C_\ell^{Ep(\eta_n)} & \dots  & C_\ell^{p(\eta_n)p(\eta_n)}
	\end{bmatrix}
    +\mathbf{N}_\ell \label{eq:covariance}
\ee
and the noise covariance is taken to be diagonal $\mathbf{N}_\ell = \mathrm{diag}\left( N_\ell^{TT}, N_\ell^{EE}, N_\ell^{p(\eta_1)p(\eta_1)}, \dots, N_\ell^{p(\eta_n)p(\eta_n)}\right)$. We use a modified version of the Boltzmann solver CAMB~\cite{Lewis:1999bs} to compute the spectra appearing in Eq.~(\ref{eq:covariance}), fixing $\ell_\mathrm{max} = 20$ for all forecasts. 


\begin{figure}[t]
\includegraphics[width = \columnwidth]{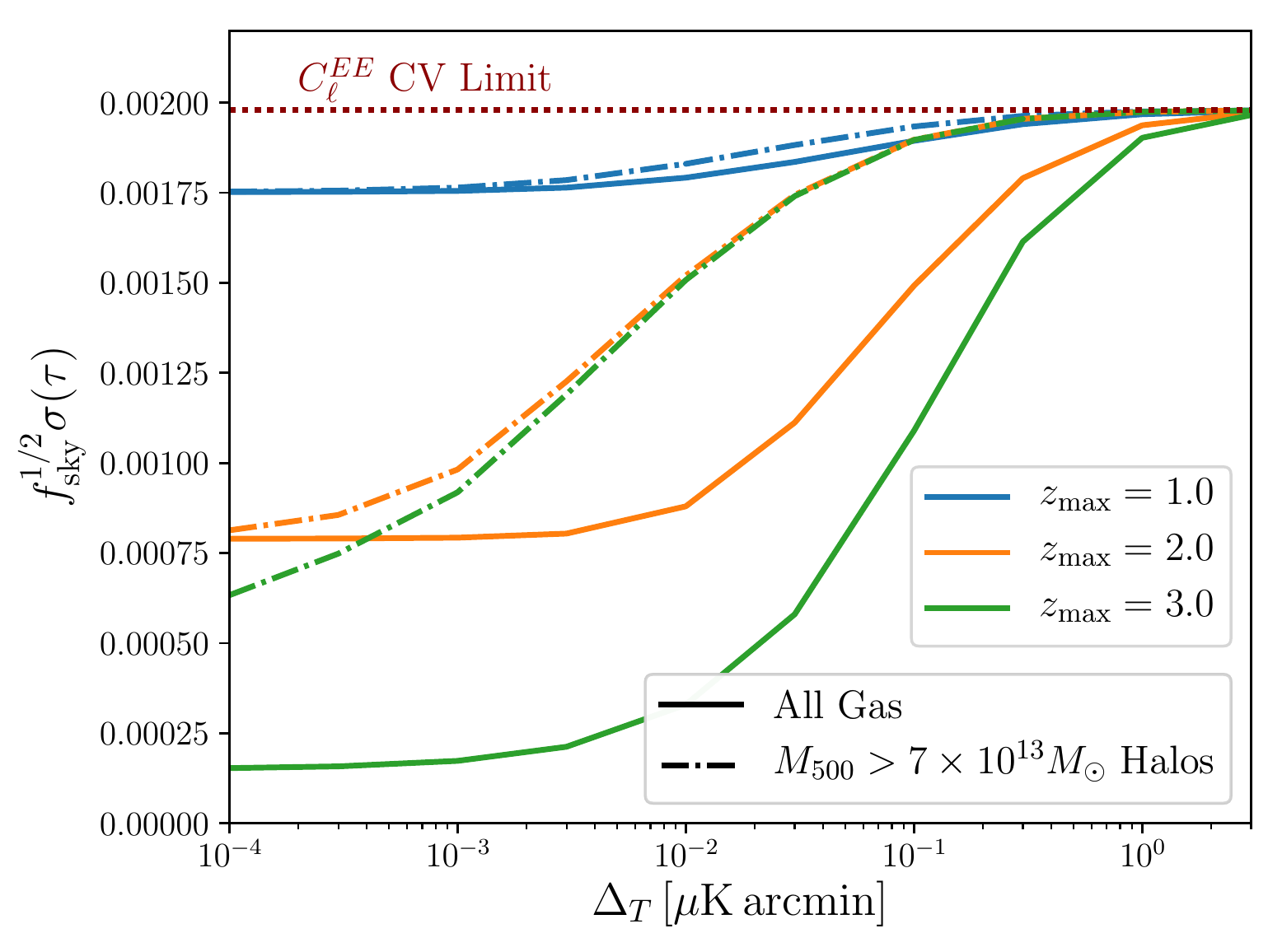}
\caption{Forecasted 1-$\sigma$ error on the mean optical depth $\tau$,  as a function of the noise level on a 1-arcminute CMB survey used to measure the pSZ effect on small angular scales, assuming vanishing noise on large scale $E$ modes $\ell\leq 20$.  Here we also assume that the effects of lensing can be completely removed. We show forecasts assuming knowledge of either all of the gas (solid) or only of halos of mass $M_{500} > 7\times 10^{13}\, M_\odot$ (dash-dot) out to three choices of maximum redshift.
} \label{fig:constraints}
\end{figure}

In Fig.~\ref{fig:constraints}, we show the forecasted 1-$\sigma$ error on the mean optical depth $\tau$, marginalized over the uncertainty on the scalar amplitude $A_s$.  The forecasts show that with sufficiently precise measurements of remote temperature quadrupoles, the error on $\tau$ can be dramatically improved over the cosmic variance limit of $E$-mode polarization alone.  

\begin{figure}[t]
\includegraphics[width = \columnwidth]{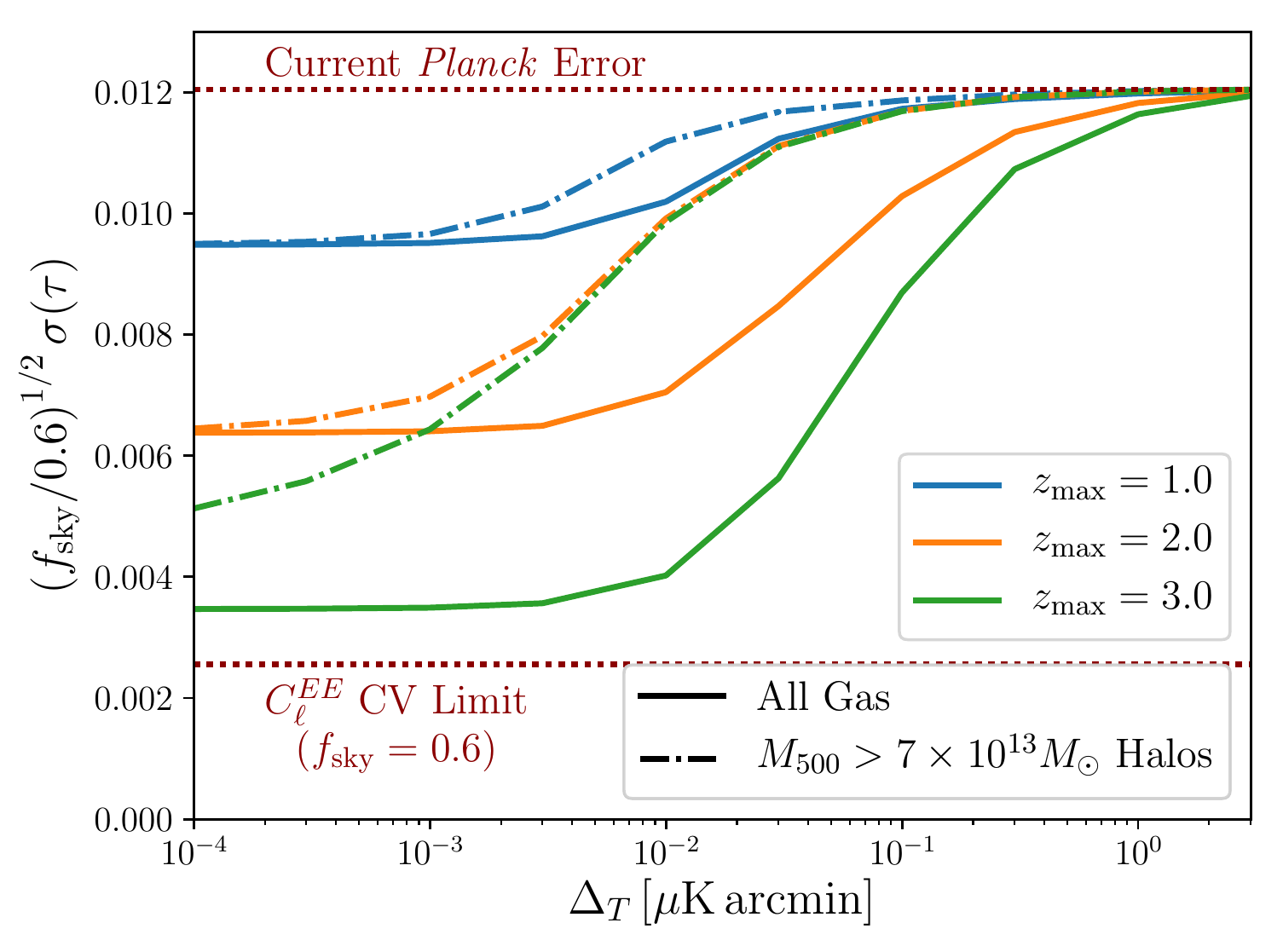}
\caption{Identical to Fig.~\ref{fig:constraints} but with \textit{Planck}-like noise on large scale $E$ modes.}\label{fig:constraints_planck}
\end{figure}

It is possible to reduce the uncertainty on $\tau$ even if errors on large scale $E$-mode polarization do not improve over the current best measurements from the \textit{Planck} satellite. To estimate this improvement, we approximated the \textit{Planck} noise by matching the systematic plus noise curve shown in Fig.~2 of Ref.~\cite{Adam:2016hgk}, and we show the result in Fig.~\ref{fig:constraints_planck}. 

\begin{figure}[t]
\includegraphics[width = \columnwidth]{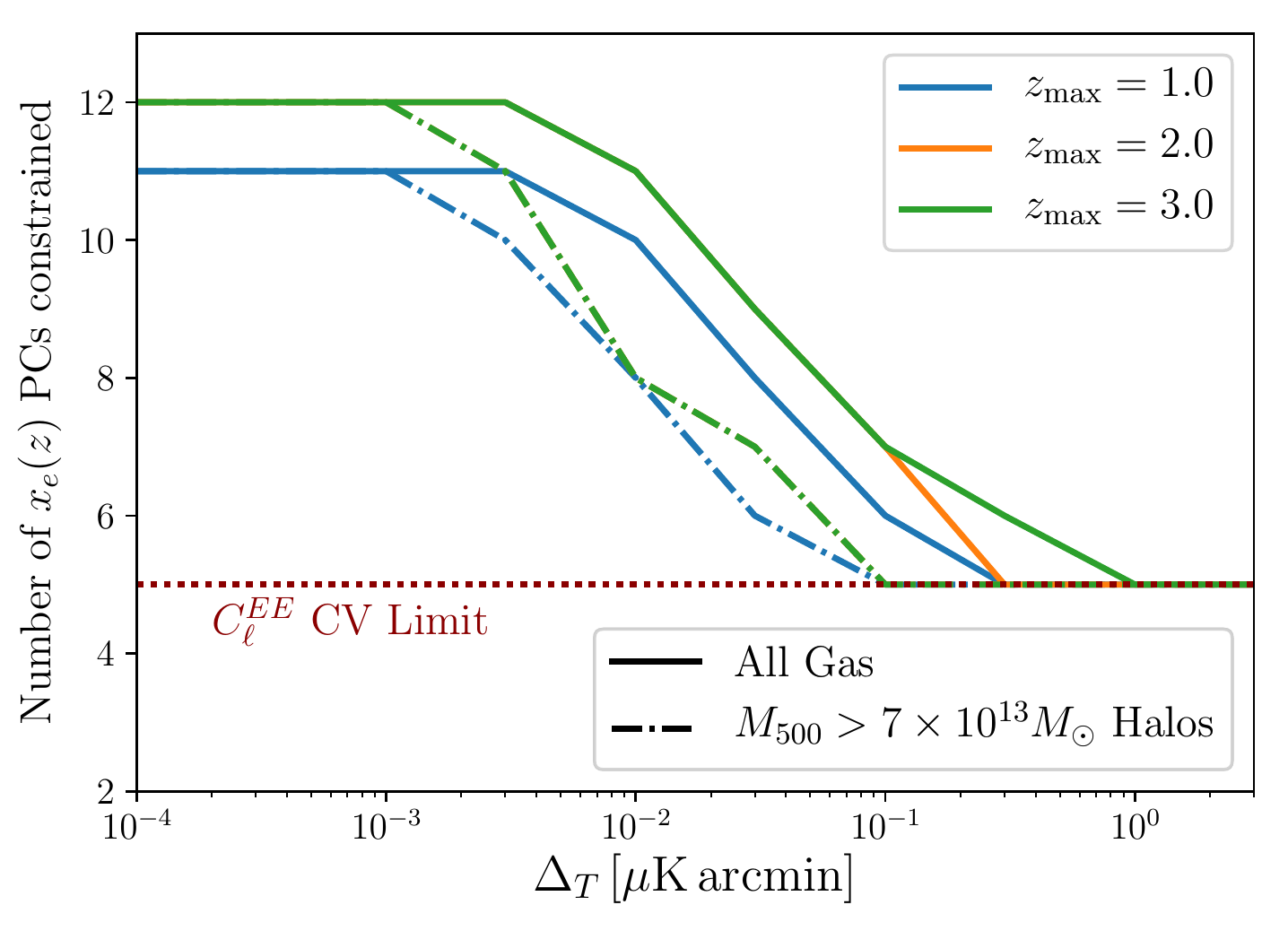}
\caption{Number of principal components of the redshift-dependent ionization fraction $x_e(z)$ which can be constrained with greater than 1-$\sigma$ precision with the same assumptions and conventions as used in Fig.~\ref{fig:constraints}.} \label{fig:nPCs}
\end{figure}

\begin{figure}[t]
\includegraphics[width = \columnwidth]{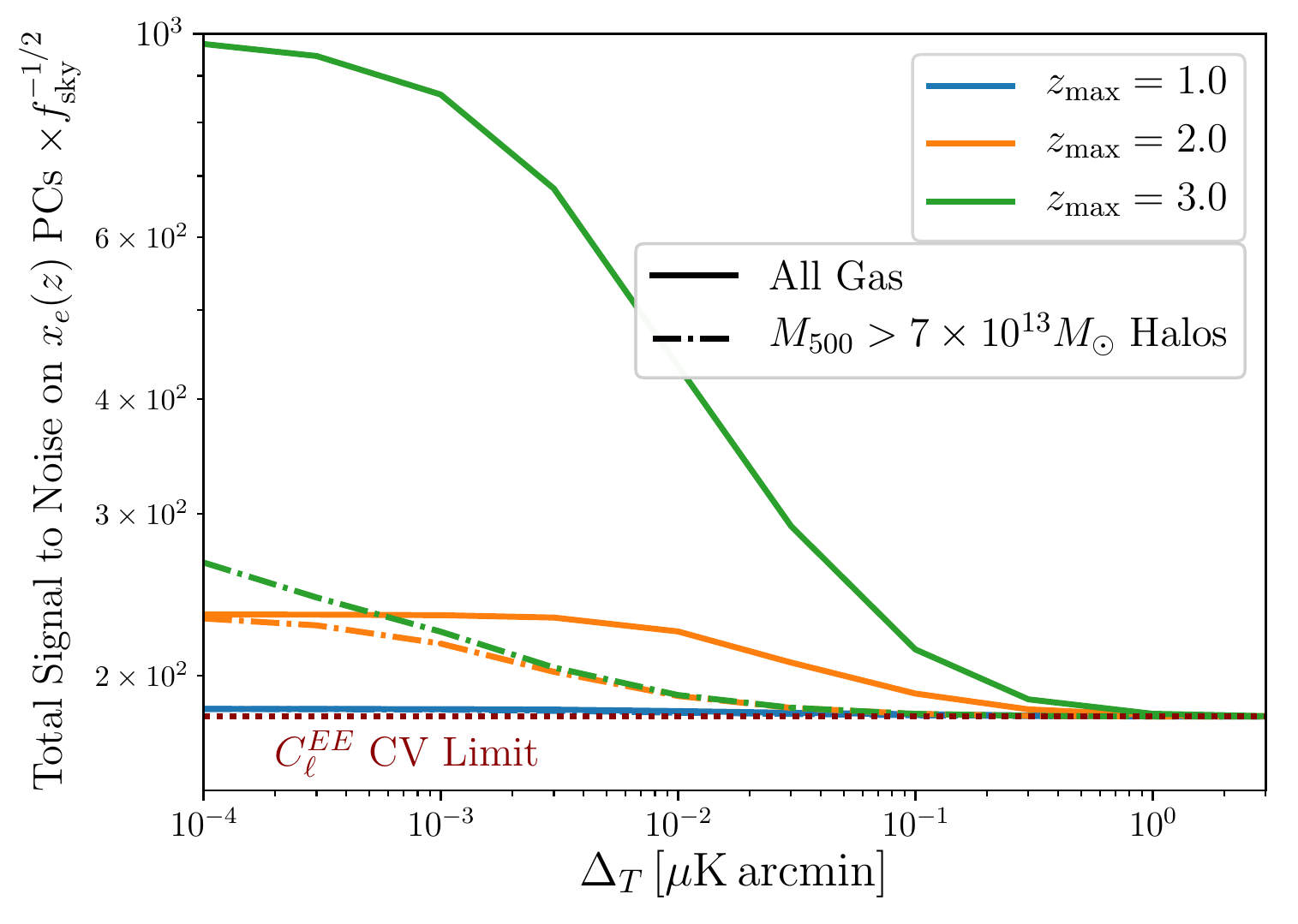}
\caption{Total signal to noise with which principal components of the redshift-dependent ionization fraction $x_e(z)$ can be measured with the same assumptions and conventions as used in Fig.~\ref{fig:constraints}.} \label{fig:snPCs}
\end{figure}

Our ability to constrain reionization goes beyond the mean optical depth.  Following the methods of Refs.~\cite{Hu:2003gh,Mortonson:2007hq}, we compute the Fisher matrix for the binned ionization fraction $x_e(z)$ using the public code from Ref.~\cite{Mortonson:2007hq}, which we then decompose into principal components.\footnote{Following Ref.~\cite{Mortonson:2007hq}, we allow the ionization fraction to vary between $z_\mathrm{min}=6$ and $z_\mathrm{max}=30$, using a bin width of $\Delta z=0.25$, and a fiducial model with $x_e(z)=0.15$ in this range.} While the $E$-mode polarization alone can constrain at most 5 principal components, or independent pieces of information, on the ionization history, we show in Fig.~\ref{fig:nPCs} that remote quadrupole measurements allow us to constrain more than twice as many principal components.  Furthermore, the signal to noise with which these principal components can be measured can be increased by more than a factor of five, as shown in Fig.~\ref{fig:snPCs}.


\section{Discussion and Conclusions}
We have shown how measurements of remote temperature quadrupoles can be used to probe the cosmic reionization history. While the method we described has the potential to greatly improve over measurements using the CMB alone, obtaining sufficiently precise measurements of the remote quadrupole field to realize this improvement will prove challenging. No experiment has yet been able to detect the pSZ effect, though the first detection should occur with upcoming CMB and galaxy surveys.  


Despite these limitations, our analysis clearly demonstrates that remote quadrupole measurements are potentially very useful for probing reionization. While we have focused here on the pSZ effect to map out remote quadrupoles, there are other methods which are capable of achieving the same goal. For example, distant temperature quadrupoles result in circular polarization of 21cm photons \cite{Hirata:2017dku}. The methods of non-linear reconstruction of large-scale modes, using their gravitational influence on the statistics of small-scale density fluctuations ~\cite{Zhu:2015zlh}, could be used to map out the large scale modes which source remote temperature quadrupoles.

The objects we assumed for our forecasts corresponded to  massive clusters that would be found internally 
in a futuristic CMB survey using the thermal Sunyaev-Zel'dovich effect. It remains to be seen whether including a catalog of lower-mass, more numerous objects, such as optically- or infrared-selected galaxies, would approach the optimistic case that we also calculated---namely, all the ionized gas to a given redshift. We also assumed that the CMB survey aimed for a uniform noise over a wide patch of sky; better results might be obtained from dedicated microwave observations of polarization in the direction of a small number of known objects~\cite{Hall:2014wna}.

We have neglected a number of effects which complicate the simple forecasts we presented here, such as the additional variance that comes from polarization generated by multiple scattering in objects, the kinematic quadrupole due to their transverse motion relative to the CMB, and lensing of background polarization by the halos hosting these objects~\cite{Cooray:2002cb,Hall:2014wna,Yasini:2016pby}. We have not included secondary sources of polarization on small scales such as from patchy reionization~\cite{Dvorkin:2009ah} or from polarized emission from galaxies.  A more careful treatment of the effect of incomplete sky coverage is warranted, since the analysis on very large angular scales is complicated by missing modes~\cite{Alizadeh:2012vy}.

We have also neglected the additional uncertainty that comes from the inference of the optical depth of the objects from an external survey.  This inference could be made, for example, by using a galaxy survey and assuming that gas is a biased tracer of large scale structure; by relating the amplitude of the thermal Sunyaev-Zel'dovich effect~\cite{Battaglia:2016xbi} or the X-ray flux~\cite{Flender:2016cjy} in each object to the optical depth of the object using a calibrated relationship; or by observing the screening of CMB fluctuations by the objects~\cite{Dvorkin:2008tf}. Scatter of these inferences about the true optical depth of the objects (due either to measurement error or intrinsic scatter in a calibrated relationship) increases the noise of the remote quadrupole reconstruction by a factor which decreases as the square root of the number of objects in the survey.  A systematic offset on these inferences would bias the reconstruction of the remote quadrupole field, leading in turn to an offset on the determination of the mean optical depth. If the method to obtain object optical depths is correlated across redshift bins, there will also appear non-diagonal terms in the noise covariance matrix appearing in Eq.~(\ref{eq:covariance}).


The determination of cosmological parameters which are key targets of future cosmological surveys, such as the sum of neutrino masses, will be limited by the uncertainty on the mean optical depth~\cite{CMBS4}. Additionally, the epoch of reionization is an exciting frontier which marks the emergence of the first luminous sources in our Universe. Understanding the physical properties of these sources which reionized the Universe requires measurements of the ionization history~\cite{Fan:2006dp,2012RPPh...75h6901P,Zaroubi:2012in}. We demonstrated that combining maps of remote temperature quadrupoles with CMB polarization on large angular scales can provide exquisite constraints on the cosmic reionization history, far better than can be achieved with the CMB alone.


\vspace{1.0em}

\noindent
\textbf {Acknowledgments }
We would like to thank Marcelo Alvarez, Colin Hill, Gil Holder, Matthew Johnson, Brayden Mon, Tyler Natoli, Connor Sheere, and David Spergel for helpful discussions. J.M. and A.v.E. were supported by the Vincent and Beatrice Tremaine Fellowship. P.D.M. acknowledges support from a Senior Kavli Fellowship at the University of Cambridge. N.B. acknowledges the support from the Lyman Spitzer Jr. Fellowship.


\appendix

\subsection*{Appendix: Remote Quadrupole Estimator}

In this appendix, we will derive the minimum variance estimator of the remote quadrupole field, given maps of the CMB polarization and of the electron density. Angular scales relevant for remote quadrupole reconstruction are sufficiently small that we can use the flat sky approximation. We aim to construct an estimator of the form
\be
\widehat{p}^{\pm}(\nhat, \eta) \sim \left[ \left(Q(\nhat)\pm i U(\nhat)\right)^{\rm obs} \delta\tau^{\rm ext}(\nhat, \eta) \right]_f
\ee
with $Q$ and $U$ the Stokes parameters of the CMB polarization and $\delta\tau^{\rm ext}(\nhat,\eta)$ the optical depth in direction $\nhat$ at a conformal look-back time $\eta$ inferred from an external survey. The subscript $f$ refers to the filtering that will be applied to the maps to minimize the variance of the estimator.  The derivation and results of this appendix are very similar to those of Ref.~\cite{Alizadeh:2012vy}, except we perform them here for the flat sky.

The observed polarization in some direction $\nhat$ contains a `primary' piece which would be present even in the absence of free electrons at the redshift of interest, a piece that results from the scattering of the temperature quadrupole at the space-time location of objects containing free electrons, and instrumental noise
\be
	\left(Q\pm iU\right)^\mathrm{obs}(\nhat) &=& \left(Q\pm iU\right)^{\rm prim}(\nhat) \nonumber \\
	&&+p^\pm(\nhat,\eta)\delta\tau(\nhat, \eta) + n^\pm(\nhat) \, .
\ee
We can express this in harmonic space using $E$- and $B$-modes for both the observed polarization and the remote quadrupole field, as in Eqs.~(\ref{eq:EBlm}) and (\ref{eq:plm}),
\be 
	\left(Q\pm iU\right)(\vl) &=& \left(E\mp iB\right)(\vl) \  e^{\mp 2i\phi_{\vl}} \nonumber \\
	p^{\pm}(\vl,\eta) &=& \left(p^E\mp ip^B\right)(\vl,\eta) \ e^{\mp 2i\phi_{\vl}} \, ,
\ee
which gives
\be 
	&&\left(E\mp iB\right)^\mathrm{obs}(\vl) = \left(E\mp iB\right)^\mathrm{prim}(\vl) + \left(n^E \mp in^B\right)(\vl) \nonumber \\
    &&\quad+\int \frac{\mathrm{d}^2\ell'}{2\pi} \left(p^E\mp ip^B\right)(\vlp,\eta) e^{\mp 2i\left(\phi_{\vlp} - \phi_{\vl}\right)}\delta\tau(\vl-\vlp,\eta).\nonumber \\
\ee

The independently inferred map of the optical depth $\delta\tau^\mathrm{ext}(\nhat, \eta)$ will typically be incomplete, containing only the contributions from objects which can be resolved in the survey being considered.  Notice that when the inferred map is a subset of the true map ($ \delta\tau = \delta\tau^\mathrm{ext} + \delta\tau^\mathrm{unobs}$ with $\langle \delta\tau^\mathrm{ext}\delta\tau^\mathrm{unobs} \rangle = 0$), we find that 
\be 
	\left\langle \delta\tau^\mathrm{ext}(\vl,\eta) \delta\tau^\mathrm{ext}(\vlp,\eta) \right\rangle &=& \left\langle \delta\tau^\mathrm{ext}(\vl,\eta) \delta\tau(\vlp,\eta) \right\rangle \\
    &=& \delta^{(2)}(\vl+\vlp)\ \rho_\ell^2(\eta)\ C_\ell^{\delta\tau\delta\tau(\eta)} \, , \nonumber
\ee
where $C_\ell^{\delta\tau\delta\tau(\eta)}$ is the power spectrum of the optical depth due to all of the free electrons at the time of interest, and we have introduced a cross-correlation coefficient $\rho_\ell^2(\eta) \leq 1$, which accounts for the incompleteness of the inferred map of $\delta\tau$.  The cross-correlation coefficient is also reduced when the inferred optical depth scatters about the true value due either to observational noise or an intrinsic scatter in the relationship between a given tracer of optical depth and its true value.

We can compute the power spectrum of the optical depth in a halo model as follows:
\be \label{eq:app_taupower}
	&& C_\ell^{\delta\tau\delta\tau(\eta(z))} =  \int_{z_\mathrm{min}}^{z_\mathrm{max}} \! \mathrm{d} z \frac{dV}{dz} \int_0^\infty \! \mathrm{d} M \frac{dn}{dM dz} \left| \tau_{M,z}(\ell) \right|^2  \\
	&& \quad + P_m^\mathrm{lin}(\ell,z) \left[ \int_{z_\mathrm{min}}^{z_\mathrm{max}} \! \mathrm{d} z \frac{dV}{dz} \int_0^\infty \! \mathrm{d} M \frac{dn}{dM dz} b(M,z) \tau_{M,z}(\ell)  \right]^2 \nonumber
\ee
where $\frac{dn}{dM dz}$ is the comoving number density of halos of mass $M$ at redshift $z$, $\tau_{M,z}(\ell)$ is the optical depth profile of a given halo, and $b(M,z)$ is the linear halo bias.  The redshift integral is performed over a bin which should be chosen to be small compared to the coherence length of the field of remote quadrupoles.  The first line represents the one-halo term, and the second line represents the two-halo term describing the clustering of objects.  For a given survey, we can calculate the coefficient $\rho_\ell^2(\eta)$ by computing Eq.~(\ref{eq:app_taupower}) with a mass cut $M_\mathrm{min}$ defined by the survey and dividing by Eq.~(\ref{eq:app_taupower}) computed with no mass cut.  See Fig.~\ref{fig:rho_halo} for an example of this correlation coefficient when only galaxy clusters are assumed to have been measured.

We can then write an estimator for $p^E$ as
\be 
	\widehat{p}^{E}(\vL) &=& N^E(\vL) \int \frac{\mathrm{d}^2{\vl}}{2\pi} \delta\tau^\mathrm{ext}(\vL-\vl)  \\
    &&\times \left(f^{Ep^E}(\vl,\vL)E^\mathrm{obs}(\vl) + f^{Bp^E}(\vl,\vL)B^\mathrm{obs}(\vl)\right)  \, , \nonumber
\ee
and we are suppressing the explicit redshift dependence.  
We demand that our estimator is unbiased
\begin{widetext}
\be 
	\widehat{p}^{E}(\vL) &=& \left\langle \widehat{p}^{E}(\vL) \right \rangle \\
    &=& N^E(\vL) \int \frac{\mathrm{d}^2{\vl}}{2\pi} \int \frac{\mathrm{d}^2{\vlp}}{2\pi} \left\langle \delta\tau^\mathrm{ext}(\vL-\vl) \delta\tau(\vl-\vlp) \right \rangle \nonumber \\
    && \times \left[ f^{Ep^E}(\vl,\vL) \left(p^E(\vl')\cos(2(\phi_{\vlp} - \phi_{\vl})) - p^B(\vl')\sin(2(\phi_{\vlp} - \phi_{\vl}))\right) \right. \nonumber \\
    && +\left. f^{Bp^E}(\vl,\vL) \left(p^E(\vl')\sin(2(\phi_{\vlp} - \phi_{\vl})) + p^B(\vl')\cos(2(\phi_{\vlp} - \phi_{\vl}))\right) \right] \, , \nonumber
\ee
which fixes the filters to be related by
\be 
	\frac{f^{Ep^E}(\vl,\vL)}{\cos(2(\phi_{\vL}-\phi_{\vl}))} = \frac{f^{Bp^E}(\vl,\vL)}{\sin(2(\phi_{\vL}-\phi_{\vl}))} \equiv f^{E}(\vl,\vL) \, ,
\ee
and the normalization to be
\be \label{eq:pp_noise}
	N^E(\vL) = \left[\int \frac{\mathrm{d}^2{\vl}}{(2\pi)^2} f^{E}(\vl,\vL) \rho_{|\vl-\vL|}^2 C_{|\vl-\vL|}^{\delta\tau\delta\tau} \right]^{-1} \, .
\ee
Next, we choose our filter to minimize the variance of our estimator.  The variance is
\be 
	&& \left\langle \widehat{p}^{E}(\vL) \widehat{p}^{E*}(\vLp) \right\rangle = \delta^{(2)}(\vL-\vLp) \left|N^E(\vL)\right|^2  \\
    && \qquad \times \int \frac{\mathrm{d}^2{\vl}}{(2\pi)^2}  \left| f^E(\vl,\vL) \right|^2 \left(C_\ell^{EE,\mathrm{obs}}\cos^2(2(\phi_{\vL}-\phi_{\vl})) + C_\ell^{BB,\mathrm{obs}}\sin^2(2(\phi_{\vL}-\phi_{\vl}))\right) \rho_{|\vl-\vL|}^2 C_{|\vl-\vL|}^{\delta\tau\delta\tau} \, , \nonumber
\ee
and the filter which minimizes the variance is
\be 
	f^E(\vl,\vL) = \left(C_\ell^{EE,\mathrm{obs}}\cos^2(2(\phi_{\vL}-\phi_{\vl})) + C_\ell^{BB,\mathrm{obs}}\sin^2(2(\phi_{\vL}-\phi_{\vl}))\right)^{-1} \, .
\ee
Using this choice of filter in Eq.~(\ref{eq:pp_noise}) then gives the variance of our estimator.

We can follow the same steps as above to derive the unbiased, minimum variance estimator for $p^B$ (which is expected to vanish in the absence of tensor fluctuations), and we find
\be 
	\widehat{p}^{B}(\vL) &=& N^B(\vL) \int \frac{\mathrm{d}^2{\vl}}{2\pi} \delta\tau^\mathrm{ext}(\vL-\vl) f^{B}(\vl,\vL) \left(-E^\mathrm{obs}(\vl)\sin(2(\phi_{\vL}-\phi_{\vl})) + B^\mathrm{obs}(\vl)\cos(2(\phi_{\vL}-\phi_{\vl}))\right)  \, , 
\ee
with the filter and normalization given by
\be 
	f^B(\vl,\vL) &=& \left(C_\ell^{EE,\mathrm{obs}}\sin^2(2(\phi_{\vL}-\phi_{\vl})) + C_\ell^{BB,\mathrm{obs}}\cos^2(2(\phi_{\vL}-\phi_{\vl}))\right)^{-1} \, , \nonumber \\
    N^B(\vL) &=& \left[\int \frac{\mathrm{d}^2{\vl}}{(2\pi)^2} f^{B}(\vl,\vL) \rho_{|\vl-\vL|}^2 C_{|\vl-\vL|}^{\delta\tau\delta\tau} \right]^{-1} \, .
\ee
\end{widetext}


\bibliography{quadrupole}

\end{document}